\documentstyle [aps,psfig]{revtex}

\begin{document}
\draft
\title{Signature for local Mixmaster dynamics in $U(1)$ symmetric cosmologies\thanks{
e-mail: berger@oakland.edu, vincent.moncrief@yale.edu}} 
\author{Beverly K. Berger}
\address{Department of Physics, Oakland University, Rochester, MI 48309 USA}
\author{Vincent Moncrief}
\address{Departments of Physics and Mathematics, Yale University, New Haven, CT
06520 USA}
\maketitle
\bigskip
\begin{abstract}
Previous studies \cite{berger98a} have provided strong support for a local, oscillatory
approach to the singularity in $U(1)$ symmetric, spatially inhomogeneous vacuum
cosmologies on $T^3 \times R$. The description of a vacuum Bianchi type IX, spatially
homogeneous Mixmaster cosmology (on $S^3 \times R$) in terms of the variables used to describe the
$U(1)$ symmetric cosmologies indicates that the oscillations in the latter are in fact
those of local Mixmaster dynamics. One of the variables of the $U(1)$ symmetric models
increases only at the end of a Mixmaster era. Such an increase therefore yields a
qualitative signature for local Mixmaster dynamics in spatially inhomogeneous cosmologies.
\end{abstract}

\section{Introduction}
In the 1960's, Belinskii, Khalatnikov, and Lifshitz (BKL) developed a formalism
to analyze the approach to the singularity in spatially inhomogneous cosmologies
\cite{belinskii69a,belinskii69b,belinskii71a,belinskii71b,belinskii82}. They
argued that, while one can construct, over any short time interval, a local Kasner
solution \cite{kasner25}, this type of solution cannot be maintained indefinitely since
terms inconsistent with it cannot remain subdominant. Rather, the generic solution will
behave as a spatially homogeneous Mixmaster cosmology
\cite{misner69} at every spatial point. For a long time, this analysis remained
controversial, primarily due to its non-rigorous nature.

Recently, in a series of papers
\cite{berger98a,berger93,berger97b,berger97e,berger98c,weaver98}, we and our collaborators
have used an analytic method, simpler than but not unrelated to that of BKL, along with
detailed numerical simulations to provide strong support for the BKL picture. In
addition, our methods have clarified how Einstein's equations control the approach to the
singularity. The essence of our method of consistent potentials (MCP) developed
originally by Grubi\u{s}i\'{c} and Moncrief
\cite{grubisic93} is to solve the velocity term dominated (VTD) equations obtained by
neglecting all terms containing spatial derivatives in Einstein's equations. We then
substitute the asymptotic form of the VTD solution (with spatially dependent parameters)
into the neglected terms. If these terms are all exponentially small, the VTD
solution is consistent. This in turn supports a conjecture that the approach to the
singularity is asymptotically VTD (AVTD) --- i.e., at almost every spatial point, the
full solution comes arbitrarily close to a VTD solution as the singularity is approached
\cite{isenberg90}. On the other hand, if one or more of the neglected terms grows
exponentially, then the VTD solution is inconsistent so that the conjectured behavior is
not AVTD. If two or more previously neglected terms will alternately always grow
for any range of VTD solution parameters, then the MCP predicts that the solution will be
oscillatory and the oscillations will persist forever. If there exists a range of
parameter values consistent with a VTD solution almost everywhere, the prediction is that
oscillations will persist only until the parameters are driven into the consistent range
\cite{berger97b}. For the MCP to predict the correct behavior,
the only dynamically significant time dependences, sufficiently close to the singularity,
must be the linear ones from the VTD solutions which appear in the arguments of the
exponential  potentials. Perhaps surprisingly then, the MCP predictions have been
verified by numerical simulation of the full Einstein equations in all models we have
studied so far \cite{berger98c}.

The previous discussion of the application of the MCP makes no appeal to an invariant
description of the system. One then wonders whether or how this picture depends on the
choice of variables and of spacetime slicing. There is also an issue of the nature of
oscillatory behavior observed in spatially inhomogeneous cosmologies \cite{berger98a}. The
BKL claim is that these models should evolve as a vacuum Bianchi type IX
(Mixmaster) universe at every spatial point. In the MCP applied to the minisuperspace
(MSS) picture of vacuum, homogeneous Bianchi type IX models, the VTD is the Kasner
solution. This solution is inconsistent with maintaining exponential smallness of the
three exponential wall terms in the MSS potential --- leading to the usual infinite
sequence of bounces off a closed triangular potential \cite{misner69}. However, we found
that vacuum $U(1)$ symmetric cosmologies on $T^3 \times R$ exhibit oscillations between
two potentials (rather than three) at every spatial point \cite{berger98a}. At the time,
we did not know if these represented ``true'' Mixmaster oscillations or something else.

In this paper, we shall describe spatially homogeneous Bianchi type IX universes on $S^3
\times R$ as $U(1)$ symmetric cosmologies with the same topology by rewriting the 
Bianchi type IX spatial metric in a coordinate frame \cite{grubisic94}. The objective of
this exercise is to see how the standard Mixmaster oscillations are described in the
$U(1)$ variables used in our previous studies. We find that the relationship
between the MSS and $U(1)$ descriptions of the Mixmaster universe is not trivial. In the
coordinate representation of the Bianchi type IX homogeneous space, one of the Killing
vectors is still manifest and is taken to play the role of the spatial $U(1)$ symmetry.
The norm of that Killing vector is the key variable in our previous description of
generic, spatially inhomogeneous $U(1)$ models. Here we find its behavior to be
dominated by the largest logarithmic scale factor (LSF) of the standard BKL description.
The bounces off one of the $U(1)$ potentials occur when the time derivative of the
dominant LSF changes sign. This is what happens in the standard MSS picture in a bounce
off any of the three potential walls. The other $U(1)$ potential causes a bounce when one
LSF loses its dominant place to another in the MSS picture. This correspondance will be
discussed extensively in this paper. 

Although numerical analysis of inhomogeneous $U(1)$ symmetric cosmologies revealed only
the two types of bounces described above, our study of the $U(1)$-MSS correspondance for
the Mixmaster universe demonstrates that a third type of bounce can occur --- at the end
of a BKL era \cite{belinskii71b} --- when the monotonically decreasing smallest LSF starts
increasing and overtakes the middle one. This yields a qualitative prediction --- not yet
observed numerically --- which can distinguish between local Mixmaster dynamics and other
oscillatory behavior.

In Section II, we shall review the MSS picture and, in Section III, the $U(1)$ variables.
Note that in this paper, we shall often use ``$U(1)$'' to refer to properties of the
$U(1)$ symmetric cosmologies.  The correspondance between the $U(1)$ models and the
Mixmaster models will be discussed in Section IV with the implications of the
correspondance given in Section V. Conclusions will be drawn in Section VI.

\section{The MSS picture of Mixmaster dynamics}
A vacuum, diagonal Bianchi type IX Mixmaster cosmology is described by the metric
\cite{misner73}
\begin{equation}
\label{mixmetric}
ds^2=-A^2B^2C^2d\tau ^2+A^2(\sigma ^1)^2+B^2(\sigma ^2)^2+C^2(\sigma^3)^2
\end{equation}
where the scale factors $A$, $B$, and $C$ are functions of $\tau$ only and 
\begin{equation}
\label{bianchi9}
d\sigma ^i={1 \over 2} \varepsilon ^i_{jk}\,\sigma ^j\wedge \sigma ^k.
\end{equation}
It is often convenient to define the LSF's $\alpha$, $\zeta$, and $\gamma$ by
\begin{equation}
\label{lsfdef}
A = e^\alpha, \quad \quad B = e^\zeta, \quad \quad C = e^\gamma \,.
\end{equation}
The $SU(2)$ symmetry (\ref{bianchi9}) may be realized in a coordinate frame by
\begin{eqnarray}
\label{sigmas}
\sigma^1 &=& \cos \phi \, d\theta + \sin \theta \sin \phi \, d\psi, \nonumber \\
\sigma^2 &=& - \sin \phi \, d\theta + \cos \phi \, \sin \theta \, d\psi, \nonumber \\
\sigma^3 &=& d\phi + \cos \theta \, d \psi
\end{eqnarray}
on $S^3$. The choice of time coordinate
\begin{equation}
\label{bkllapse}
N \, d\tau = ABC \, d\tau = dt,
\end{equation}
where $t$ is comoving proper time, is that used by BKL.

An alternate, but equivalent description, was given by Misner \cite{misner69} in terms of
the MSS variables $\Omega$, the logarithmic volume, and $\beta_\pm$, the anisotropic
shears, where
\begin{eqnarray}
\label{convertbkl}
\alpha &=& \Omega - 2 \beta_+, \nonumber \\
\zeta &=& \Omega + \beta_+ + \sqrt{3} \beta_-, \nonumber \\
\gamma & =& \Omega +  \beta_+ - \sqrt{3} \beta_-.
\end{eqnarray}
This variable choice conveniently allows Einstein's equations to be obtained by variation
of the superhamiltonian 
\begin{equation}
\label{generalh0}
\int_{S^3} \,NH  =\int_{S^3} \, {N \over {\sqrt{g}}} \left[ \left( \pi^{ij} \pi_{ij} - {1
\over 2} \pi^2 \right) - g {}^3\/R \right]
\end{equation}
where $g$ is the determinant of the spatial metric $g_{ij}$ with conjugate momenta
$\pi^{ij}$ while ${}^3\!R$ is the scalar curvature of $g_{ij}$ and $H = 0$ is the
Hamiltonian constraint. With $p_\Omega$, $p_\pm$ canonically conjugate to
$\Omega$, $\beta_\pm$ respectively, and the time coordinate choice $N = \sqrt{g}/\sin
\theta $, Eq.~(\ref{generalh0}) becomes
\begin{equation}
\label{mixh0}
2H = 0 = - p_\Omega^2 + p_+^2 + p_-^2 + V_{IX}(\beta_\pm, \Omega)
\end{equation}
for the MSS potential
\begin{equation}
\label{msspot}
V_{IX} = e^{4\alpha} + e^{4\zeta} + e^{4\gamma} - 2 e^{2(\alpha+\zeta)} - 2 e^{2(\zeta +
\gamma)} - 2 e^{2(\gamma + \alpha)}.
\end{equation}
When $V_{IX}$ is exponentially small, the metric is locally (neglecting the topology) the
Kasner solution. Generically, in the evolution toward the singularity, one of the first
three terms on the right hand side of Eq.~(\ref{msspot}) --- the one associated with
the largest LSF --- will grow. A ``bounce'' off the exponential potential will change the
sign of the time derivative of the dominant LSF changing one Kasner solution into another.
Conservation of momentum in the bounce was first used by BKL to relate the two asymptotic
Kasner solutions. In terms of the LSF's, Einstein's evolution equations for the Mixmaster
model are (for overdot indicating $ d/d\tau$)
\begin{eqnarray}
\label{mixeqs}
\ddot \alpha &=& {1 \over 2} \left[ \left( e^{2 \zeta} - e^{2 \gamma} \right)^2 - e^{4
\alpha} \right], \nonumber \\
\ddot \zeta &=& {1 \over 2} \left[ \left( e^{2 \gamma} - e^{2 \alpha} \right)^2 - e^{4
\zeta} \right], \nonumber \\
\ddot \gamma &=& {1 \over 2} \left[ \left( e^{2 \alpha} - e^{2 \zeta} \right)^2 - e^{4
\gamma} \right].
\end{eqnarray}
If the
initial Kasner solution is characterized by a given
$\{\dot \alpha, \dot \zeta, \dot \gamma \}$ and the dominant potential term is
$e^{4\alpha}$, then from Eqs.~(\ref{mixeqs}) the final asymptotic Kasner solution with $\{
\dot \alpha',\, \dot \zeta',\, \dot \gamma' \}$ will have $\dot \alpha' = - \dot \alpha$,
$\dot \zeta' = \dot \zeta + 2 \dot \alpha$, $\dot \gamma' = \dot \gamma + 2 \dot \alpha$.
Various schemes exist to encode this information
\cite{belinskii71b,chernoff83,berger96c}. If, initially, $\dot \alpha > \dot  \zeta > 
\dot \gamma$, then initially $\dot \alpha > 0$ while $\dot \zeta < 0$, $\dot \gamma < 0$
since the (collapsing) Kasner solution generically has only one expanding direction. This
means that $\alpha$ increases while the others decrease. After a typical bounce, the above
bounce rules yield $\dot \alpha' < 0$, $\dot \zeta' > 0$ while $\dot \gamma$ has
increased but is still negative. In subsequent bounces, $\alpha$ and $\zeta$ will
continue to oscillate while $\gamma$ decreases monotonically but with ever decreasing $|
\dot \gamma |$. (Each Kasner regime between the bounces is called an epoch.) Eventually,
$\dot \gamma$ will change sign. This is called the end of an era. If
$\dot \alpha > \dot \zeta > \dot \gamma$ prior to this era ending bounce, we find that,
asymptotically, $\dot \gamma' > \dot \zeta' > \dot \alpha'$ after the bounce.
Subsequently, $\zeta$ and $\gamma$ oscillate while $\alpha$ decreases monotonically. This
sequence of eras apparently \cite{rendall97b,weaver99b,ringstrom99} persists
indefinitely. Part of a typical Mixmaster evolution is shown in Fig.~1.

\section{$U(1)$ symmetric cosmologies on $S^3 \times R$}
$U(1)$ symmetric cosmologies on $S^3 \times R$ are described by the metric
\cite{moncrief86}
\begin{equation}
\label{u1metric}
ds^2 = e^{-2 \varphi} \{- \tilde N^2 \, d\tau^2 + \tilde g_{ab} (dx^a + \tilde N^a \,
d\tau)(dx^b + \tilde N^b \,d\tau) \} + e^{2 \varphi} (d\psi + \cos \theta d\phi + \beta_a
\, dx^a + \beta_0 \, d\tau)^2
\end{equation}
where the symmetry direction is $\psi$, and the other spatial directions are $\{x^a\} =
\{\theta,\phi\}$. The metric variables are assumed to be functions of $\theta$, $\phi$,
and $\tau$. The norm of the Killing field is $e^\varphi$, $\beta_a$ are the ``twists'',
$e^{2 \Lambda}$ is the determinant of the 2-metric $\tilde g_{ab} = e^\Lambda e_{ab}$ and
$e_{ab}$ is parametrized by $x$ and $z$ via
\begin{equation}
\label{eab}
e_{ab}={1 \over 2}\left[
{\matrix{{e^{2z}+e^{-2z}(1+x)^2}&{e^{2z}+e^{-2z}(x^2-1)}\cr
{e^{2z}+e^{-2z}(x^2-1)}&{e^{2z}+e^{-2z}(1-x)^2}\cr }} \right] \ .
\end{equation}
Note that the metric $(\ref{u1metric})$ differs from that for $T^3$ spatial topology given
in \cite{berger98a}. It is convenient to make a canonical transformation from the twists
and their conjugate momenta $e^a$ to the twist potential $\omega$ and its conjugate
momentum $r$. It is also convenient to define the spacetime slicing by zero shift and
lapse $\tilde N \sin \theta = \sqrt{\tilde g} = e^\Lambda$ where $\tilde g$ is the
determinant of the 2-metric $\tilde g_{ab}$. Einstein's equations are found by variation
of 
\begin{eqnarray}
\label{Hu1}
H &=& \int_{S^3} \,{\cal H} \nonumber \\
&=& \int_{S^3} {1 \over {\sin \theta}} \left[ \left( {1 \over 8}p_z^2+{1 \over 2}
e^{4z}p_x^2+{1 \over 8}p^2+{1 \over 2}e^{4\varphi }r^2-{1 \over 2}p_\Lambda
^2  \right) \right. \nonumber \\
&& + \left\{  \left( {e^\Lambda e^{ab}} \right) ,_{ab}-
\left( {e^\Lambda e^{ab}}
\right) ,_a\Lambda ,_b+e^\Lambda  \right. \left[  \left( {e^{-2z}}
\right) ,_u x,_v- \left( {e^{-2z}} \right) ,_v x,_u \right] \nonumber \\
&& \left. \left. +2e^\Lambda e^{ab}\varphi ,_a\varphi ,_b+{1 \over 2}
e^\Lambda e^{-4\varphi }e^{ab}\omega ,_a\omega ,_b \right\} \right] 
\end{eqnarray}
where ${\cal H} = 0$ is the Hamiltonian constraint and the overall trigonometric factor
comes from $N/\sqrt{g}$.

An excellent approximation to the behavior seen in numerical simulations of these models
(albeit on $T^3$ rather than $S^3$) is that at each spatial point a Kasner-like phase
characterized by \cite{berger98a}
\begin{eqnarray}
\label{vtdsoln}
\varphi &=& - v_\varphi \tau , \quad \quad 
p = -4 v_\varphi,  \quad \quad
z = - v_z \tau, \nonumber \\
p_z &=& -4 v_z,  \quad \quad
\Lambda = -v_\Lambda \tau, \quad \quad
p_\Lambda = v_\Lambda, \nonumber \\
\omega &=& \omega_0, \quad \quad
r = r_0, \quad \quad
x = x_0, \quad \quad
p_x = p_x^0 
\end{eqnarray}
(where $v_\varphi$, $v_z$, $v_\Lambda$, $\omega_0$, $r_0$, $x_0$, and $p_x^0$ are
functions of $\theta$ and $\phi$ but independent of $\tau$) is followed by a ``bounce''
off one of the three potentials 
\begin{eqnarray}
\label{u1pots}
V_1 &=& {1 \over 2} r^2 e^{4 \varphi}, \nonumber \\
V_2 &=& {1 \over 2}\,e^\Lambda e^{-4 \varphi} e^{ab} \omega,_a \omega,_b \, , \nonumber \\
V_3 &=& {1 \over 2} p_x^2 e^{4z}.
\end{eqnarray}
In Eqs.~(\ref{u1pots}), the potentials are defined to be identical to those in
\cite{berger98a}. (It is possible to incorporate the prefactor $N/\sqrt{g}$ from
Eq.~(\ref{Hu1}) into the potentials. However, this factor depends on topology. In
Section V we shall see that only the (subdominant) logarithm of this factor would appear
in the variables of interest and would not affect the qualitative behavior.) If one
assumes incident and final asymptotic solutions of the form (\ref{vtdsoln}) at a given
spatial point and that $r$, $\omega$, $x$, and $p_x$ are constant in time through the
bounce, the remaining momenta $p$, $p_z$, and $p_\Lambda$ at that point will change
during a bounce off one of the  potentials according to the following rules:

(1) For a bounce off $V_1$, $p \to -p$ with the other momenta conserved.

(2) For a bounce off $V_2$,
\begin{eqnarray}
\label{v2rules}
p &\to& -p - p_z - 2 p_\Lambda, \nonumber \\
p_z &\to& {1 \over 2} (-2 p + p_z -2 p_\Lambda), \nonumber \\
p_\Lambda &\to& {1 \over 4} (2 p + p_z + 6 p_\Lambda). 
\end{eqnarray}
These rules are obtained by recognizing that, for $z$ large and negative, $e^{ab}$ is
dominated by $e^{-2z}$ so that $V_2 \approx K e^{\Lambda -2z -4 \varphi}$ where $K$ is
approximately constant in time. Then $\Lambda-2z-4\varphi$ defines the direction (in a
local MSS) in which the conjugate momentum changes sign during a bounce with
momenta in the plane orthogonal to that direction conserved. 

(3) For a bounce off $V_3$, $p_z \to -p_z$ with the other momenta conserved.

(4) In all cases, the momenta must satisfy an asymptotic form of the Hamiltonian
constraint (\ref{Hu1}) given by
\begin{equation}
\label{h0lim}
{\cal H}_{lim} = - {1 \over 2} p_\Lambda^2 + {1 \over 8} p^2 + {1 \over 8} p_z^2 = 0.
\end{equation}
Eq.~(\ref{h0lim}) is obtained from Eq.~(\ref{Hu1}) by assuming that all terms containing
exponential potentials are exponentially small.

In numerical simulations performed so far \cite{berger98a}, the dominant local behavior is
oscillation of $\varphi$ due to bounces off $V_1$ and $V_2$. A bounce off $V_3$ has been
seen only if $p_z > 0$ initially. Note that the rule (\ref{v2rules}) for a bounce off
$V_2$ becomes $p \to -p$ if $|p/p_z| << 1$ implying $p_z \approx 2 p_\Lambda$ (from
Eq.~(\ref{h0lim})). Our simulations provide strong support for BKL's claim that the
approach to the singularity in generic cosmological spacetimes is oscillatory. The
remaining question then is whether or not this oscillatory behavior represents local
Mixmaster dynamics.

\section{The Mixmaster universe as a $U(1)$ symmetric cosmology}
By writing the Mixmaster metric (\ref{mixmetric}) (denoted $g^{IX}$ in this section) in
the coordinate frame on $S^3$ given by Eq.~(\ref{sigmas}) and comparing it to the $U(1)$
metric (\ref{u1metric}) (denoted $g^U$ in this section), each $U(1)$ variable $\varphi$,
$z$, $\Lambda$, $\omega$, or $x$ and its conjugate momentum may be expressed in terms of
the BKL scale factors $A$, $B$, and $C$ and their time derivatives. The explicit Killing
direction $\psi$ will be identified with the $U(1)$ symmetry direction $x^3$ so that
$e^{2 \varphi} = g^{IX}_{\psi \psi}$ or
\begin{equation}
\label{e2phi}
e^{2\varphi} = A^2 \sin^2 \theta \sin^2 \phi + B^2 \sin^2 \theta \cos^2 \phi + C^2 \cos^2
\theta.
\end{equation}
The twists $\beta_\theta$ and $\beta_\phi$ in $g^U$ are determined from
$e^{2 \varphi} \beta_\theta = g^{IX}_{\theta \psi}$ and $e^{2 \varphi} (\beta_\phi+ \cos
\theta) = g^{IX}_{\phi \psi}$ so that 
\begin{eqnarray}
\label{betas}
\beta_\theta &=& {{(A^2 - B^2) \cos \phi \sin \theta \sin \phi} \over {C^2 \cos^2 \theta
+ \sin^2 \theta (B^2 \cos^2 \phi + A^2 \sin^2 \phi)}}, \nonumber \\
\beta_\phi &=& {{C^2 \cos \theta} \over {C^2 \cos^2 \theta
+ \sin^2 \theta (B^2 \cos^2 \phi + A^2 \sin^2 \phi)}}-\cos \theta.
\end{eqnarray}
Then the $U(1)$ model 2-metric (here denoted as $\tilde g_{ab} = e^\Lambda e_{ab}$) is
found from
\begin{eqnarray}
\label{2metric}
g^{IX}_{\theta \theta} &=& e^{-2\varphi}\tilde g_{\theta \theta} + e^{2 \varphi}
\beta_\theta^2, \nonumber \\
g^{IX}_{\theta \phi} &=& e^{-2\varphi}\tilde g_{\theta \phi} + e^{2 \varphi} \beta_\theta
(\beta_\phi+\cos \theta), \nonumber \\
g^{IX}_{\phi \phi} &=& e^{-2\varphi}\tilde g_{\phi \phi} + e^{2 \varphi}
(\beta_\phi+\cos \theta)^2.
\end{eqnarray}
Comparison of Eqs.~(\ref{mixmetric}) and (\ref{u1metric}) with Eqs.~(\ref{e2phi}) and
(\ref{betas}), computation of the determinant of $\tilde g$, and the
expression of $e_{ab}$ in terms of $x$ and $z$ (see Eq.~(\ref{eab})) leads to
\begin{equation}
\label{lamequiv}
e^\Lambda = ABC \sin \theta \sqrt{C^2 \cos^2 \theta + \sin^2 \theta (B^2 \cos^2 \phi +
A^2 \sin^2 \phi)},
\end{equation}
\begin{equation}
\label{zequiv}
e^{2z} = {{4 ABC \sin \theta \sqrt{C^2 \cos^2 \theta + \sin^2 \theta (B^2 \cos^2 \phi +
A^2 \sin^2 \phi)}} \over {A^2 B^2 + A^2 C^2 + B^2 C^2 - A^2 B^2 \cos (2 \theta) + (A^2 -
B^2) C^2 \cos [2(\theta -\phi)]}},
\end{equation}
\begin{equation}
\label{xequiv}
x =  {{A^2 B^2 +[B^2 C^2 +A^2(C^2-B^2)] \cos(2\theta) +(A^2-B^2)C^2 \cos(2\phi)} \over 
{A^2 B^2 + A^2 C^2 + B^2 C^2 - A^2 B^2 \cos (2 \theta) + (A^2 -
B^2) C^2 \cos [2(\theta -\phi)]}} \ .
\end{equation}

We note that 
\begin{equation}
\label{detgab}
\sqrt{\tilde g} = e^\Lambda = ABC \sin \theta \,
e^\varphi.
\end{equation}
Our gauge condition in \cite{berger98a} (modified for $S^3$) is $\tilde N = e^\Lambda /
\sin
\theta$. But this is the ``2-dimensional lapse.'' The 3-dimensional lapse is $N = \tilde
N e^{-\varphi}$ so that 
\begin{equation}
\label{bkllapse}
N = ABC
\end{equation}
which is just the usual BKL lapse condition \cite{belinskii71b}. This means that the
time variable $\tau$ chosen in $U(1)$ models is the same as the usual BKL time
$\tau$.

It is now necessary to consider the canonical transformation which replaces $\beta_a$ and
its conjugate momentum $e^a$ subject to the constraint \cite{moncrief86}
\begin{equation}
\label{twistconstraint}
e^a,_a = 0
\end{equation}
by $\omega$ and its canonically conjugate momentum $r$ where
\begin{equation}
\label{wdef}
e^a = \varepsilon^{ab}\omega,_b
\end{equation}
for $\varepsilon^{ab} = - \varepsilon^{ba}$. In these variables, $e^a$ may be found from
the Einstein equation for the time derivative of $\beta_a$ \cite{moncrief86}:
\begin{equation}
\label{betadot}
e^a = {{\sqrt{\tilde g}} \over {\tilde N}} e^{4 \varphi} \tilde g^{ab} \beta_b,_\tau.
\end{equation}
We find
\begin{eqnarray}
\label{es}
e^\theta &=& {{2 \sin^2 \theta} \over N} ( BC \dot A \sin \phi \cos \phi - AC \dot B
\sin \phi \cos \phi) = \omega,_\phi , \nonumber \\
e^\phi &=& {{2 \sin \theta \cos \theta} \over N} (AB \dot C - BC \dot A \sin^2 \phi -
AC \dot B \cos^2 \phi) = -\omega,_\theta
\end{eqnarray}
to give
\begin{equation}
\label{weval}
\omega = {{\sin^2 \theta } \over N} (-AB \dot C + BC \dot A \sin^2 \phi + AC \dot B
\cos^2 \phi) + k(\tau)
\end{equation}
where $k(\tau)$ is an as yet undertermined function of time.

The momentum conjugate to $\omega$ is defined by an integrability condition
\cite{moncrief86} so that 
\begin{eqnarray}
\label{rdef}
r &=& \beta_\theta,_\phi - \beta_\phi,_\theta \, + \sin \theta \nonumber \\
  & = & e^{-4\varphi} \left \{ (A^2 - B^2) \sin^3 \theta (B^2 \cos^2 \phi - A^2 \sin^2
\phi) \right. \nonumber \\
  & & + \left.  C^2 \sin \theta [A^2 + B^2 - C^2 + \sin^2 \theta (C^2 - A^2 \cos^2
\phi - B^2 \sin^2 \phi)] \right \}  .
\end{eqnarray}
It is easily checked that $\int_{S^2} r = 4 \pi$ as is required. From the $U(1)$
symmetric model Hamiltonian (\ref{Hu1}), we obtain
\begin{equation}
\label{wdot}
\omega,_\tau = r e^{4 \varphi} { {\tilde N} \over {\sqrt{\tilde g}}}
\end{equation}
so that
\begin{equation}
\label{kdef}
k(\tau) = - \left ( {{\dot A} \over A} + {{\dot B} \over B} \right ) + k_0
\end{equation}
where $k_0$ is a constant.

The other momenta --- $p$, $p_z$, $p_\Lambda$, $p_x$ --- may also be found from the
equations of motion --- for $\varphi,_\tau$, $z,_\tau$, $\Lambda,_\tau$, $x,_\tau$
respectively and are given for completeness in the Appendix. 

Given all the variables, it becomes a simple matter to construct all the terms in the
$U(1)$ Hamiltonian (\ref{Hu1}). In particular, we shall consider
\begin{eqnarray}
\label{u1v1mix}
V_1 & =& {1 \over 2} r^2 e^{4 \varphi} \nonumber \\
     &=&\left\{ {(A^2-B^2)\sin ^3\theta \left( {B^2\cos ^2\phi 
-A^2\sin ^2\phi } \right)} \right. \nonumber \\
& & \left. {+C^2\sin \theta \left[ {A^2+B^2-C^2+\sin ^2\theta 
\left( {C^2-A^2\cos ^2\phi -B^2\sin ^2\phi } \right)} \right]}
\right\}^2 \nonumber \\
& &  /\left[ {2\left( {C^2\cos ^2\theta +B^2\cos ^2\phi 
\;\sin ^2\theta +A^2\sin ^2\phi \;\sin ^2\theta } \right)} \right]^2,
\end{eqnarray}
\begin{eqnarray}
\label{u1v2mix}
V_2 &=&{1 \over 2} e^{\Lambda - 4 \varphi} e^{ab} \omega,_a \omega,_b \nonumber \\
&=&\sin ^3\theta \left\{ {\left[ {A^2B^2\sin ^2\theta 
+C^2\cos ^2\theta \left( {A^2\cos ^2\phi +B^2\sin ^2\phi } \right)}
\right]\sin ^2(2\phi )\left( {B\dot A-A\dot B} \right)^2} \right.\nonumber \\
& &  +4\cos ^2\theta \left( {B^2\cos ^2\phi +A^2\sin ^2\phi } \right)
\left[ {AC\dot B\cos ^2\phi +B\left( {C\dot A\sin ^2\phi -A\dot C}
\right)} \right]^2 \nonumber \\
& &  +4\left( {A^2-B^2} \right)C\cos ^2\theta \cos \phi 
\sin \phi \sin (2\phi )\left( {A\dot B-B\dot A} \right)\left[ {-AC\dot
B\cos ^2\phi } \right. \nonumber \\
& &  \left. {\left. {+B\left( {A\dot C-C\dot A\sin ^2\phi } 
\right)} \right]} \right\}/\left\{ {2A^2B^2\left[ {C^2\cos ^2\theta +\sin
^2\theta \left( {B^2\cos ^2\phi +A^2\sin ^2\phi } \right)} \right]^2}
\right\},
\end{eqnarray}
\begin{eqnarray}
\label{u1vzmix}
V_3 &=& {1 \over 2} p_x^2 e^{4 z} \nonumber \\
&=&2 \sin^2 \theta \left\{ {AC\cos (\theta -\phi )\left[ 
{-C^2\cos \theta \sin (\theta -\phi )+A^2\sin ^2\theta \sin \phi }
\right]\dot B} \right. \nonumber \\
& & +B^3\cos \phi \sin ^2\theta \sin (\theta -\phi )
\left( {C\dot A-A\dot C} \right)+B\cos (\theta -\phi )\left[ {C^3\dot
A\cos \theta \sin (\theta -\phi )} \right. \nonumber \\
& & \left. {\left. {-A^3\dot C\sin ^2\theta \sin \phi } 
\right]} \right\}^2/\left\{ {\left\{ {A^2\left[ {C^2\cos ^2(\theta -\phi
)+B^2\sin ^2\theta } \right]} \right.} \right. \nonumber \\
& &  \left. {\left. {+B^2C^2\sin ^2(\theta -\phi )}
\right\}^2\left[ {C^2\cos ^2\theta +\sin ^2\theta \left( {B^2\cos ^2\phi
+A^2\sin ^2\phi } \right)} \right]} \right\} ,
\end{eqnarray}
and the factor
\begin{equation}
\label{u1fmix}
f = e^{\Lambda - 2z} ={1 \over 2}\left\{ {A^2\left[ {C^2\cos ^2(\theta -\phi )
+B^2\sin ^2\theta } \right]+B^2C^2\sin ^2(\theta -\phi )} \right\} .
\end{equation}

\section{Significance of the Mixmaster-$U(1)$ Correspondance}
As was described in \cite{berger98a}, the key to the dynamics of the $U(1)$ models is the
behavior of the variable $\varphi$. In the approach to the singularity, one BKL scale
factor always dominates. The nature of Mixmaster dynamics is that a bounce occurs
when one scale factor is $\approx 1$ (with the corresponding LSF $\approx 0$). The
other scale factors are $<<1$ with the corresponding LSF's large and negative. We
shall thus refer to scale factors and functions thereof as ``order unity'' ($\approx
1$) or ``exponentially small'' ($<<1$). During the Kasner epochs (away from the
bounces), all the scale factors are exponentially small. Assume that during a Kasner epoch
$A > B > C$ and define $b = B/A$, $c = C/A$, and $\tilde c = C/B$ so that $b$, $c$, and
$\tilde c$ are $< 1$. The following discussion can be changed to reflect any other
ordering of the scale factors by performing the appropriate permutation while keeping in
mind that different scale factors will be associated with different $(\theta, \phi)$
dependences. From Eq.~(\ref{e2phi}), we see that
$\varphi$ is dominated by the largest LSF. For our chosen ordering,
\begin{equation}
\label{phiorder}
\varphi =\alpha +{1 \over 2}\ln \left[ {\sin ^2\theta \sin ^2\phi 
+b^2\left( {\sin ^2\theta \cos ^2\phi +\tilde c^2\cos ^2\theta } \right)}
\right].
\end{equation}
Since the ({\it ersatz}) spatial dependence appears in the logarithm, the spatially
homogeneous behavior dominates at any particular spatial point. Note also that if $b$
and $\tilde c$ are exponentially small, 
\begin{equation}
\label{philimit}
\varphi \approx \alpha +{1 \over 2} \ln (\sin^2 \theta \sin ^2 \phi) \approx \alpha
\end{equation}
at a generic spatial point. Figure 2 shows a superposition of $\varphi$ constructed from
Eq.~(\ref{e2phi}) on the same graph as Fig.~1. The scale factors are obtained from a
numerical simulation of the Mixmaster ODE's \cite{berger96c}. Note that
$\varphi$ changes sign in two ways. If $\varphi,_\tau > 0$ ($v_\varphi < 0$), $\varphi$
will ``bounce'' when the scale factor which dominates it bounces. This is the usual
Mixmaster bounce. However,
$\varphi,_\tau$ also changes sign, after the usual bounce, when the now-increasing scale
factor ($B$ in our case) becomes as large as the decreasing dominant scale factor ($A$ in
our case). Evaluating
$V_1$ from Eq.~(\ref{u1v1mix}) when $A > B > C$, we find, for $b$, $c$, and $\tilde c$
exponentially small, that 
\begin{equation}
\label{v1order}
\ln V_1 \approx 4 \alpha + {1 \over 2} \ln (\sin^2 \theta ),
\end{equation}
i.e.~bounces off $V_1$ correspond to the usual Mixmaster bounces. This is not
surprising since $V_1$ comes from terms in the spatial scalar curvature. (It appears in
the ``kinetic'' part of the Hamiltonian constraint because the canonical transformation
$(e^a,\,\beta_a) \to (r, \, \omega)$ interchanges momentum and configuration variables.) 

If we assume the Kasner time dependence---$\dot A = k_A A$, $\dot B = k_B B$, and $\dot
C = k_C C$ (recalling that $\dot {} = d/d\tau$ while the Kasner solution is $A = A_0
e^{k_A \tau}$, for $k_A$ a constant, etc.), we find in evaluating
$V_2$ from Eq.~(\ref{u1v2mix}) that the dominant scale factor $A$ does not appear so that
(for $\tilde c$ exponentially small),
\begin{equation}
\label{v2order}
\ln V_2 \approx 2(\zeta - \alpha) + \ln{{(k_A-k_B)^2 \sin^2 \theta \sin^2(2\phi)} \over
{2(\sin^2 \phi + b^2 \cos^2 \phi)}}
\end{equation}
implying that $V_2$ is exponentially small unless $b = e^{\zeta - \alpha}$ is order unity
rather than exponentially small. But $b = 1$ corresponds to the change in the dominant
scale factor from $A$ to $B$ as $A$ decreases while $B$ increases after the standard
Mixmaster bounce. Thus bounces off $V_2$ in terms of the $U(1)$ variables correspond to
the change in the identity of the dominant scale factor as seen in the behavior of
$\varphi$ in Fig.~2. Figure 3 illustrates the interplay among $V_1$, $V_2$, and $\varphi$
which is very similar to that seen in generic $U(1)$ models (see Figs.~2--5 in
\cite{berger98a}).

An effect not mentioned in our previous studies of generic $U(1)$ models appears when we
consider the variable $z$.  For $A > B > C$, and $b$ and $c$ exponentially small but
$\tilde c$ not necessarily $<<1$, we find from Eq.~(\ref{zequiv}) that
\begin{equation}
\label{zorder}
z \approx  \gamma - \zeta  + \ln {{\sin^2 \theta |\sin \phi|} \over {\sin^2 \theta +
\tilde c^2 \cos^2(\theta - \phi)}}
\end{equation}
---i.e.~$z$ decreases monotonically (is dominated by the most negative LSF) unless
the era ends and $C$ grows to be as large as $B$ so that $\tilde c \approx 1$. The
behavior of $z$ is shown in Fig.~4 for the same simulation as in the other figures.
Figure 5 shows $\alpha$, $\zeta$, $\gamma$, and
$z$ in the region of a bounce which changes the sign of $v_z$. The bounce rules
(\ref{v2rules}) imply that such a sign change is possible if $v_\varphi >
v_z/2+v_\Lambda/4$.  Of course, $v_z < 0$ will eventually cause a bounce off
$V_3$.

This rise of $z$, qualitatively indicating the end of a Mixmaster era, has not been
observed in previously published simulations of $U(1)$ models \cite{berger98a}. However,
in Fig.~6a of \cite{berger98a}, it is seen that $v_z$ changes during bounces off $V_2$. It
is also seen that the magnitude of $v_z$ decreases at such bounces just as in Fig.~4.
Presumably, if the simulations could be followed to larger $\tau$, $z$ would eventually
begin to increase as $v_z$ changes sign. Fig.~6a of \cite{berger98a} also shows $\Lambda$
decreasing monotonically with decreasing magnitude of the slope $v_\Lambda$. The bounce
rules (\ref{v2rules}) do not permit the sign of $v_\Lambda$ to change since
Eq.~(\ref{h0lim}) may be used to show that $6\,p_\Lambda + 2\, p \, + \, p_z$ is always
$>0$.

In \cite{berger97e,berger98a}, it was argued that all terms in ${\cal H}$ (from
Eq.~(\ref{Hu1})) containing spatial derivatives, except for $V_2$,
were proportional to
$f = e^{\Lambda - 2z}$. If $A > B > C$ in a Mixmaster model, $f \approx {\textstyle {1
\over 2}} A^4\, \left[ c^2 \cos^2 (\theta -\phi) + b^2 \sin^2 \theta + b^2 c^2
\sin^2(\theta -
\phi) \right] $ which is always exponentially small since $A \approx 1$ and $b,c << 1$.
Of course, $e^\Lambda = A^4 bc \sin \theta \sqrt{c^2 \cos^2 \theta + \sin^2 \theta (b^2
\cos^2 \phi + \sin^2 \phi)}$ which is always exponentially small. Thus, we have shown that
all the observed generic $U(1)$ behavior described in \cite{berger98a} is characteristic
of Mixmaster behavior expressed in $U(1)$ variables.

In order for this to be correct, we also expect the remaining variables $\omega$ and $x$
and all the momenta to be of order unity at all times so that, generically, they may be
regarded as approximately constant in time. This assumption underlies the MCP.  We find 
that $\omega$ is independent of the scale factors with
\begin{equation}
\label{wlim}
\omega = -k_A-k_B + \sin^2 \theta (-k_C + k_B \cos^2 \phi + k_A \sin^2 \phi).
\end{equation}
We consider the approximate forms of the remaining variables for the case
$A > B > C$ with $b$, $c$, and $\tilde c$ exponentially small. We find
\begin{eqnarray}
\label{xlim}
&&\quad x \, \approx 1, \quad p \approx 4 k_A \sin \theta , \quad p_z \approx 
{2(k_C-k_B)} \sin \theta ,
  \nonumber \\ 
&&p_\Lambda \approx -(2k_A+k_B+k_C) \sin \theta, \quad r \approx -{1 \over {\sin \theta
\sin^2 \phi}}, \quad p_x \approx {{(k_B-k_C)
\cos(\theta-\phi)} \over { \sin \phi}}.
\end{eqnarray}

\section{Conclusions}
Describing a Mixmaster universe as a $U(1)$ symmetric cosmology has yielded significant
insight into the observed behavior of generic $U(1)$ models. The oscillations in
$\varphi$ observed in the generic models are precisely what would be expected from local
Mixmaster dynamics. Perhaps of greater interest, a new effect---the signature for the end
of an era as seen in the rise of
$z$---has been predicted. Observation of this effect in longer running simulations of
generic $U(1)$ models or in models with an era change at some spatial point early in
the simulation would provide strong evidence that the observed oscillations are indeed due
to local Mixmaster dynamics. Because the rise in
$z$ is expected to be dramatic, any numerical uncertainties (see the discussion of
numerics in \cite{berger98a}) should be irrelevant. Simulations to look for this effect
are in progress.

One question which remains in our larger program to analyze the approach to the
singularity is the dependence of the MCP picture and our interpretation of the results on
our choice of spacetime slicing and variables. While the $U(1)$ description of a Bianchi
type IX Mixmaster model uses the same slicing so that no conclusions may be drawn on that
issue, we clearly have two completely different descriptions of the same Mixmaster model.
Both descriptions are characterized by intermittent Kasner (or VTD) behavior with bounces
off exponential potentials changing one Kasner solution into another. However, the three
traditional BKL LSF's are replaced in the $U(1)$ description by a
single variable $\varphi$ whose oscillations follow the change from one Kasner epoch
to the next. The $U(1)$ variable $z$ clearly signals an era change---which is not
obvious when following the usual BKL LSF's. In the study of generic collapse, we
conjecture that, in any convenient variables which allow detection of local VTD behavior,
the MCP will predict correctly whether or not the model is AVTD. In addition, it should
also be possible to identify local Mixmaster dynamics through
the comparison of the departures from AVTD behavior with the description of a homogeneous
Mixmaster universe in the same variables.

\section*{Appendix}
Using the evolutions equations for the $U(1)$ variables, we find the corresponding
momenta to be
\begin{equation}
\label{pdef}
p =4\sin \theta \frac{A\,{\sin^2 (\theta )}\,{\sin^2 (\phi )}\,\dot A + 
    B\,{\cos^2 (\phi )}\,{\sin^2 (\theta )}\,\dot B + 
  C\,{\cos^2 (\theta )}\,\dot C}{
     {C}^2\,{\cos^2 (\theta )} + 
    {\sin^2 (\theta )}\,\left( {B}^2\,
        {\cos^2 (\phi )} + {A}^2\,{\sin^2 (\phi )}
       \right) },
\end{equation}
\begin{eqnarray}
\label{pzdefa}
p_z&=&\left\{ {\left\{ {-2\sin \theta \left[ {(B^2\sin ^2\theta 
\cos ^2\phi +C^2\cos ^2\theta )\left( {A^2B^2+A^2C^2-B^2C^2} \right.} \right.} \right.}
\right. \nonumber \\
&&  -A^2B^2\cos (2\theta )\left. {+(A^2+B^2)C^2\cos [2(\theta -\phi )]} \right) \nonumber
\\
&&
 \left. {\left. {-4A^2B^2C^2\sin ^2\theta \sin ^2(\theta -\phi )\sin ^2\phi } 
\right]} \right\}{{\dot A} \over A} \nonumber \\
 && +\left\{ {2\sin \theta \left[ {(C^2\cos ^2\theta +A^2\sin ^2\theta \sin ^2\phi )
\left( {-A^2B^2+A^2C^2-B^2C^2} \right.} \right.} \right. \nonumber \\
&&  \left. {+A^2B^2\cos (2\theta )+(A^2+B^2)C^2\cos [2(\theta -\phi )]} \right) \nonumber
\\
&&  \left. {\left. {+4A^2B^2C^2\sin ^2\theta \cos ^2(\theta -\phi )\cos ^2\phi } \right]} 
\right\}{{\dot B} \over B} \nonumber \\
&&  +\left\{ {8A^2B^2C^2} \right.\cos ^2\theta +\left( {-A^2B^2+A^2C^2+B^2C^2+A^2B^2\cos
(2\theta )} \right. \nonumber \\
&&  \left. {+(A^2-B^2)C^2\cos [2(\theta -\phi )]} \right)\left[ {-A^2-B^2} \right.
\nonumber \\
&& \left. \left. \left.+(A^2-B^2)\cos (2\phi ) \right] \right\} \sin ^3\theta  {{\dot
C}  \over C} \right\}/\left\{ {\left[ {A^2B^2+A^2C^2+B^2C^2} \right.} \right. \nonumber \\
&&  -A^2B^2\cos (2\theta )+(A^2-B^2)C^2\cos [2(\theta -\phi )]\left[ {C^2\cos ^2\theta } 
\right. \nonumber \\
&&  \left. {\left. {+\sin ^2\theta (B^2\cos ^2\phi +A^2\sin ^2\phi )} \right]} \right\},
\end{eqnarray}
\begin{eqnarray}
\label{pxdef}
p_x &=& \sin \theta \left\{ \csc \theta \,\left[ A\,C\,
       \cos (\theta  - \phi )\,
       \left( - {C}^2\,\cos \theta \,
            \sin (\theta  - \phi )  \right. 
\right. \right. \nonumber \\
&& + 
    \left.     {A}^2\,{\sin^2 \theta }\,\sin \phi 
         \right) \,\dot B + 
      {B}^3\,\cos \phi \,{\sin^2 \theta }\,
       \sin (\theta  - \phi )\,
       \left( C\,\dot A - A\,\dot C \right) \nonumber \\
&& 
   \left. \left.    + B\,\cos (\theta  - \phi )\,
       \left( {C}^3\,\cos \theta \,
          \sin (\theta  - \phi )\,\dot A - 
         {A}^3\,{\sin^2 \theta }\,\sin \phi \,\dot C
         \right)  \right]  \right\} \nonumber \\
&& / \left\{{A\,B\,C\,
    \left[ {C}^2\,{\cos^2 \theta } + 
      {\sin^2 \theta }\,
       \left( {B}^2\,{\cos^2 \phi } + 
         {A}^2\,{\sin^2  \phi }\right)  \right] } \right\},
\end{eqnarray}
\begin{eqnarray}
\label{plamdef}
p_\Lambda &=& -\sin \theta \left[ 2\,A\,{B}^2\,C\,
       {\cos^2 \phi }\,{\sin^2 \theta }\,\dot B + 
      A\,C\,
       \left( {C}^2\,{\cos^2 \theta } \right. \right. \nonumber \\
&& + \left.
         {A}^2\,{\sin^2 \theta }\,{\sin^2 \phi }
         \right) \,\dot B + 
      {B}^3\,{\cos^2 \phi }\,{\sin^2 \theta }\,
       \left( C\,\dot A + A\,\dot C \right) 
       + B\,\left( {C}^3\,
          {\cos^2 \theta }\,\dot A \right. \nonumber \\
&&+ 
      \left.   \left.  2\,{A}^2\,C\,{\sin^2 \theta }\,
          {\sin^2 \phi }\,\dot A + 
         2\,A\,{C}^2\,{\cos^2 \theta }\,
          \dot C + {A}^3\,{\sin^2 \theta }\,
          {\sin^2 \phi }\,\dot C \right)  \right] \nonumber \\
&& / \left\{{A\,B\,C\,
    \left[ {C}^2\,{\cos^2 \theta } + 
      {\sin^2 \theta }\,
       \left( {B}^2\,{\cos^2 \phi } + 
         {A}^2\,{\sin^2 \phi } \right)  \right] } \right\}  .
\end{eqnarray}

\section*{Acknowledgments}
We would like to thank the Institute for Theoretical Physics at the University of
California / Santa Barbara for hospitality.  BKB would like to thank the Institute for
Geophysics and Planetary Physics of Lawrence Livermore National Laboratory and the Max
Planck Institut f\"{u}r Gravitationsphysik for hospitality. This work was supported in
part by National Science Foundation Grants PHY9732629, PHY9800103, PHY9973666, and
PHY9407194. Some of the computations discussed here were performed at the National Center
for Supercomputing Applications at the University of Illinois.

\vfill
\eject

\section*{Figure Captions}
\bigskip
Figure 1. The LSF's for a portion of a typical Mixmaster trajectory. Note that, prior to
$|\Omega| \approx 2$, $\alpha$ and $\gamma$ oscillate while $\zeta$ decreases
monotonically. This ``era'' ends when $\zeta$ starts to increase. Now $\gamma$ decreases
monotonically while $\alpha$ and $\zeta$ oscillate. It is clear that another era has
ended (but off the scale) near $|\Omega| \approx 3.5$ since $\gamma$ is seen to
oscillate again. 

\bigskip

Figure 2. The $U(1)$ symmetric model variable $\varphi$ superposed on the Mixmaster
trajectory segment of Fig.~1. It is clear that $\varphi$ always tracks the largest LSF and
that $\varphi,_\tau$ changes sign both at the original ``bounces'' of the LSF's and when a
different scale factor becomes the largest.

\bigskip

Figure 3. The $U(1)$ symmetric model variable $\varphi$ and potentials $V_1$ and $V_2$
for part of the Mixmaster trajectory segment of Fig.~1.

\bigskip

Figure 4.  The $U(1)$ symmetric model variable $z$ superposed on the Mixmaster trajectory
segment of Fig.~1. Note that $z$ decreases monotonically except at the end of an era when
it increases.

\bigskip

Figure 5. Details of the rise of $z$ in Fig.~4. (a) A standard LSF bounce (in $\zeta$)
causes $z,_\tau$ to become more negative. This is a bounce off $V_1$. (b) The next bounce
in $\varphi$, off $V_2$ occurs when $\alpha = \zeta$ and causes $z,_\tau$ to become
positive. (c) The next bounce (in $\alpha$) causes $z,_\tau$ to increase. (d) $z$ reaches
a maximum when the now increasing smallest LSF $\gamma$ equals the middle one $\zeta$.

\begin{figure}[bth]
\begin{center}
\makebox[4in]{\psfig{file=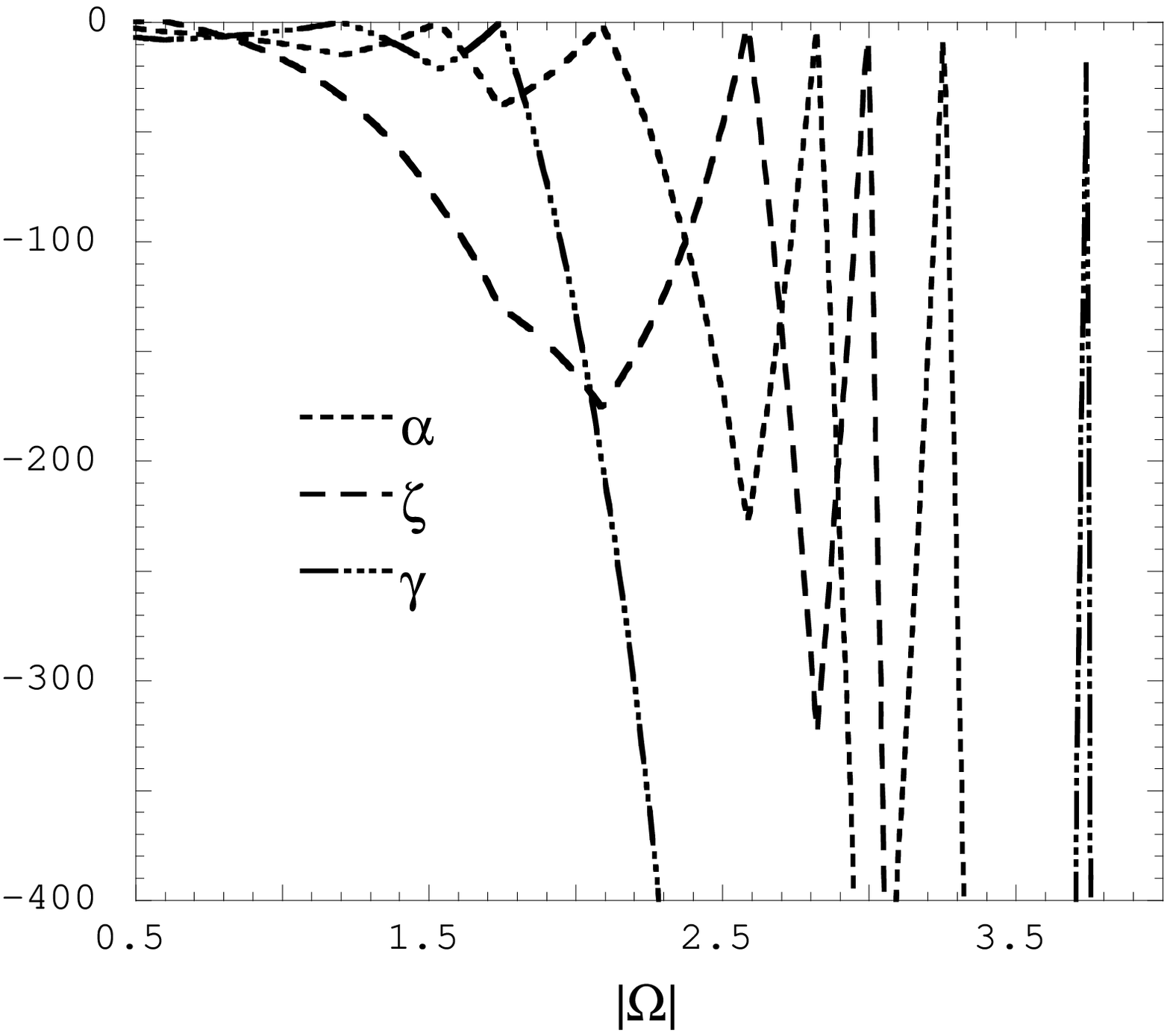,width=3.5in}}
\caption{}
\end{center}
\end{figure}
\begin{figure}[bth]
\begin{center}
\makebox[4in]{\psfig{file=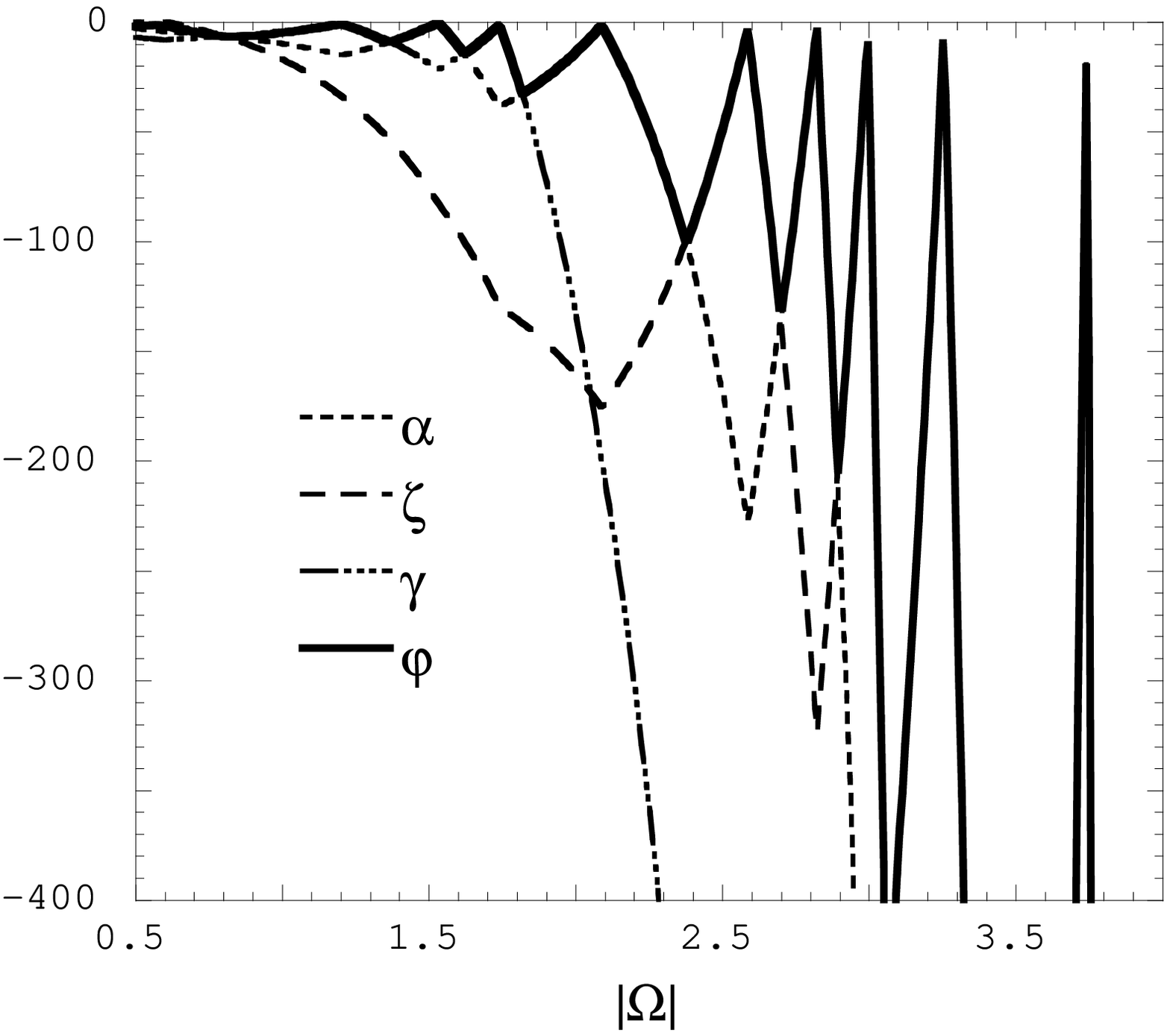,width=3.0in}}
\caption{}
\end{center}
\end{figure}
\begin{figure}[bth]
\begin{center}
\makebox[4in]{\psfig{file=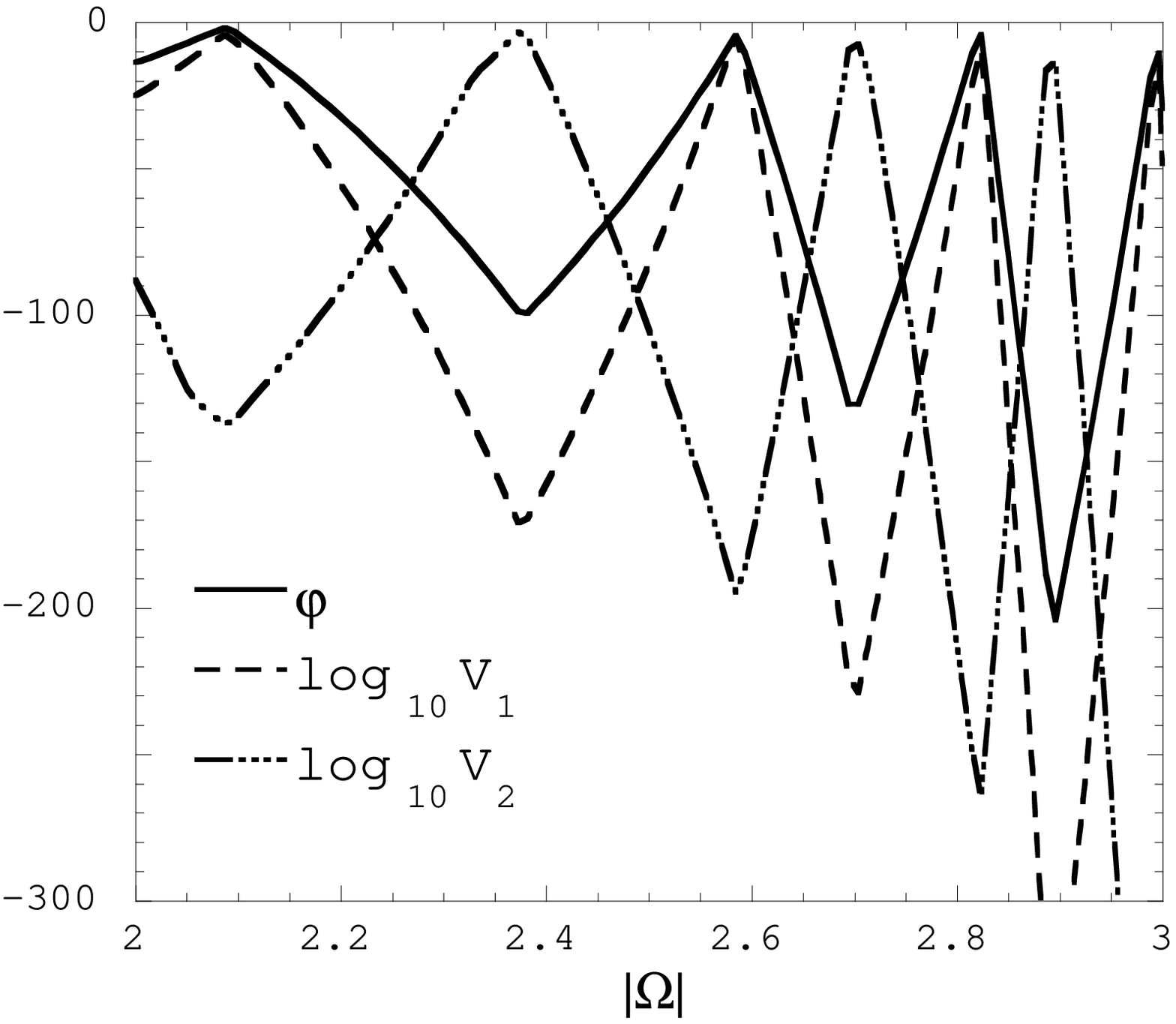,width=3.5in}}
\caption{}
\end{center}
\end{figure}
\begin{figure}[bth]
\begin{center}
\makebox[4in]{\psfig{file=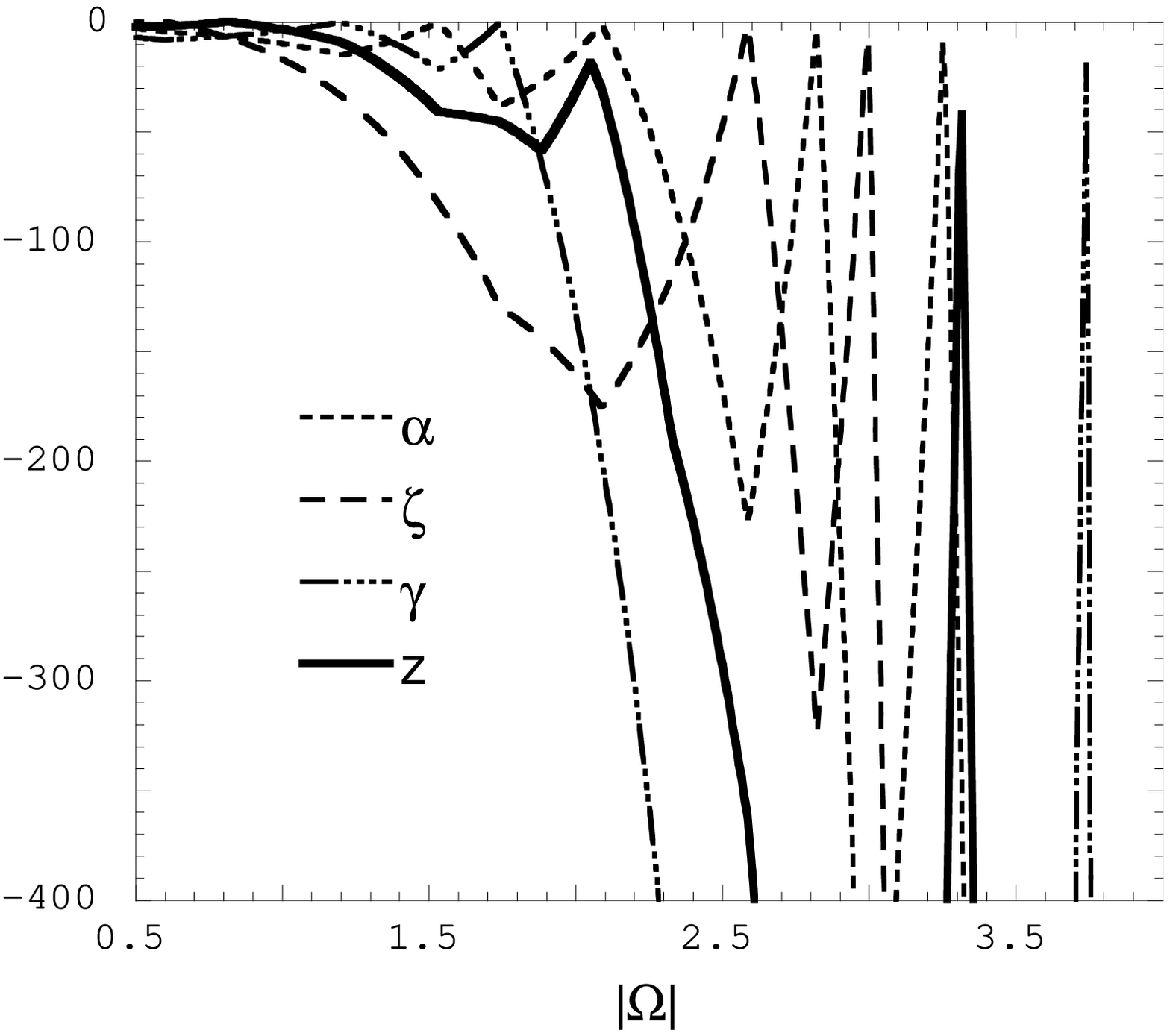,width=3.0in}}
\caption{}
\end{center}
\end{figure}
\begin{figure}[bth]
\begin{center}
\makebox[4in]{\psfig{file=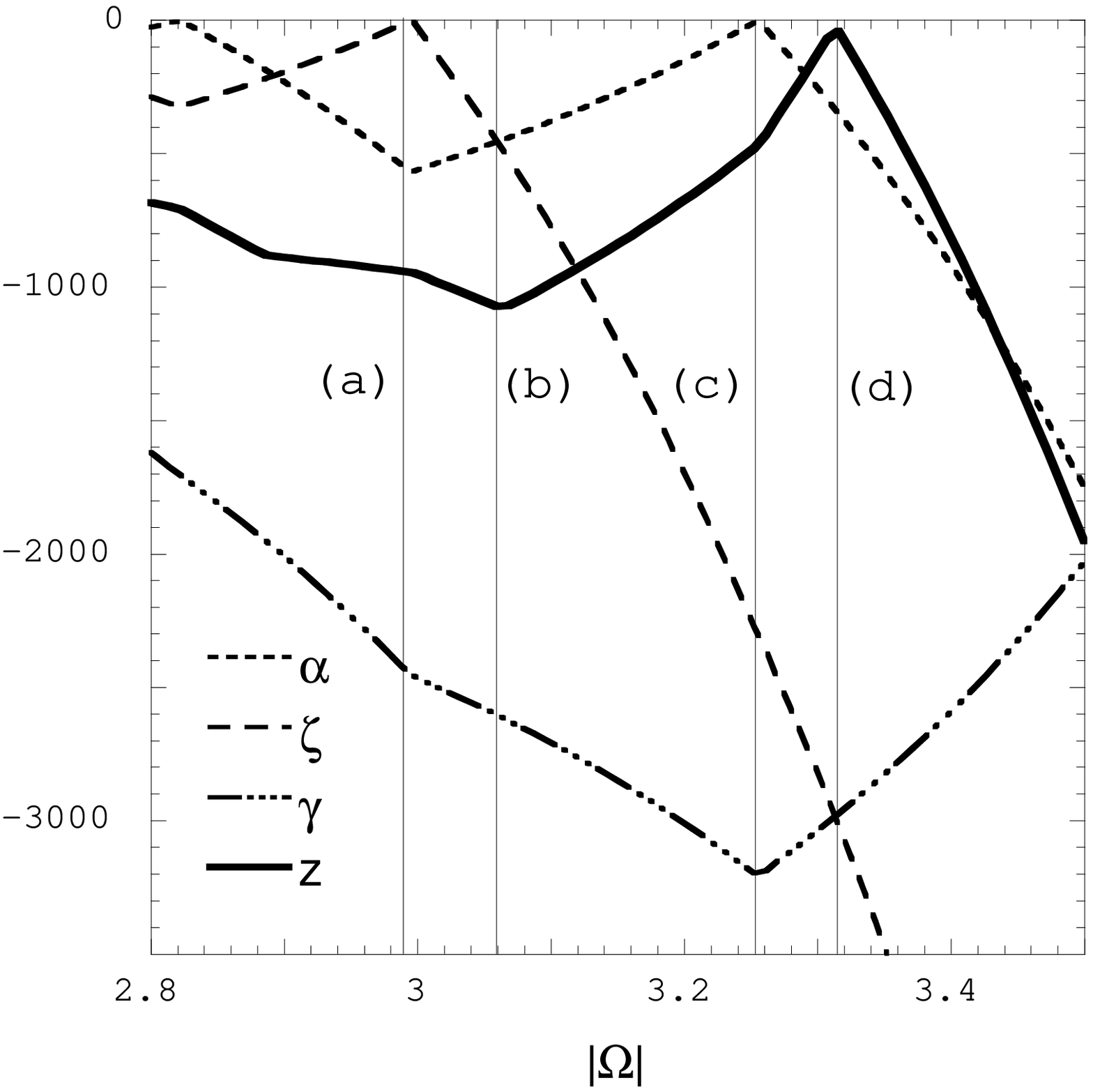,width=3.5in}}
\caption{}
\end{center}
\end{figure}

\end{document}